\documentclass[12pt]{article}
\usepackage[english]{babel}
\usepackage{amssymb,slashed,latexsym,amsmath,multirow,color}
\usepackage{slashed}
\pdfoutput=1
\usepackage[font=footnotesize,labelsep=newline,labelfont=sc,justification=centering,position=top]{caption}

\numberwithin{equation}{section}

\usepackage{hyperref}

\newsavebox{\uuunit}
\sbox{\uuunit}
    {\setlength{\unitlength}{0.825em}
     \begin{picture}(0.6,0.7)
        \thinlines
        \put(0,0){\line(1,0){0.5}}
        \put(0.15,0){\line(0,1){0.7}}
        \put(0.35,0){\line(0,1){0.8}}
       \multiput(0.3,0.8)(-0.04,-0.02){12}{\rule{0.5pt}{0.5pt}}
     \end {picture}}

\def\be{\begin{equation}}
\def\ee{\end{equation}}
\def\ba{\begin{array}}
\def\ea{\end{array}}
\def\bea{\begin{eqnarray}}
\def\eea{\end{eqnarray}}
\def\bd{\begin{displaymath}}
\def\ed{\end{displaymath}}

\begin{document}
\begin{center}
\Large{The small-scale structure of quantum spacetime}
\end{center}
\begin{center}
C. D. Burton
\end{center}
\begin{center}
\textsl{George and Cynthia Woods Mitchell Institute for Fundamental Physics and\\
Astronomy, Texas A$\&$M University, College Station, TX 77843, USA}
\end{center}
\begin{center}
\textsl{email: chris.burton@tamu.edu}
\end{center}
\
\
\begin{center}
ABSTRACT
\end{center}
\begin{sloppypar}
Planck-scale quantum spacetime undergoes probabilistic local curvature fluctuations whose distributions cannot explicitly depend on position otherwise vacuum's small-scale quantum structure would fail to be statistically homogeneous.  Since the collection of fluctuations is a many-body system, the natural explanation for their position-independent statistics is that they are in equilibrium with each other and distributed at maximum entropy.  Consequently, their probability distributions obey the laws of statistical physics which enforces small-scale smoothness, prevents the homogeneity-violating diffusion found in any free quantum system, and maintains decoherence.  Their entropy, calculated using the explicitly-constructed phase space of the Riemann whose statistics are derived using a background-independent graviton exchange ensemble, is proportional to the Einstein-Hilbert action evaluated on the macroscopic expected geometry and includes a small, positive cosmological constant.  Entropy maximization yields quantum spacetime's Ehrenfest equations of motion which are identical to Einstein's expectation-valued field equations.  This background-independent dynamical formulation reveals curvature fluctuation entropy as a source of expansion and raises the possibility that matter's zero-point energy problem, which is action-based and not energy shift invariant, may not be a problem after all. 
\end{sloppypar} 
\newpage

\newpage

\tableofcontents


\newpage

\section{Introduction}\label{intro}
\begin{sloppypar}
The cosmological constant problem has plagued physics for a number of years.  Its various aspects and potential solution categories are described in Weinberg \cite{weinberg} and subsequent progress has been reviewed in \cite{dereview1}, \cite{DEreview}.  Why is there such a large discrepancy between matter's zero point energy and the measured value of $\Lambda$?  What type of matter field would produce a stress tensor $\Lambda g_{\mu\nu}$?  Why is the cosmological constant so small?  Seemingly unrelated is the problem of decoherence in quantum gravity \cite{decohere1}, \cite{decohere}.  Why does the quantum metric remain in a ``localized'' state giving us the large-scale ``classical'' universe in which we live?  What prevents geometry's microscopic structures from quantum mechanically diffusing out of ``localized'' states?  Many-body interaction equilibrium relates these problems and provides a new path towards their solution.               
\end{sloppypar}
\begin{sloppypar}
Our current understanding of Planck-scale quantum spacetime is that it is dominated by configuration fluctuations. When described probabilistically, the entropy content of these fluctuations will have an accompanying pressure so surely this is the source of the cosmological constant or is its dominant contributor.  The principle purpose of this paper is to demonstrate that this is indeed the case by deriving directly from first principles of many-body statistical physics the Ehrenfest equations of motion $\langle G_{\mu\nu} \rangle+\Lambda \langle g_{\mu\nu}\rangle =0$ and the accompanying $\Lambda$.   
\end{sloppypar}
\begin{sloppypar}
The statistical analysis of a probabilistic, locally fluctuating geometry begins in section \ref{manybody} where the quantum Riemann is expanded as fluctuations about its expectation value.  It is argued that many-body interaction equilibrium is what prevents quantum diffusion out of the Riemann's statistically homogeneous probability distribution, enforces vacuum's small-scale statistical homogeneity, and acts as the ``environment'' producing a decoherent quantum metric.  Because the quantum Riemann obeys a canonical distribution, large $\partial^2 g$ fluctuations are exponentially suppressed which enforces small-scale smoothness enabling the use of Riemann normal metric expansions.  The Riemann tensor's phase space is explicitly constructed in section \ref{riemphase} and its entropy is explicitly formulated.  
\end{sloppypar}
\begin{sloppypar}
Section \ref{gravex} demonstrates that if a quantum field theory locally creates and destroys spacetime volume in discreet quanta, the Ricci scalar will fluctuate.  Section \ref{curvestat} uses a background-independent graviton exchange ensemble to determine the Ricci scalar's statistics which are required to find the Riemann's entropy.  Section \ref{tsgeo} calculates the entropy of the Riemann which yields quantum spacetime's Einstein-Hilbert entropy per volume whose maximization produces the Ehrenfest equations of motion $\langle G_{\mu\nu} \rangle+\Lambda \langle g_{\mu\nu}\rangle=0$.  Fluctuation entropy's small, positive $\Lambda$ results in macroscopic deSitter spacetime when no matter fields are present.  Jacobson \cite{jacobson} has previously derived Einstein's equations as thermodynamic equations of state and a thermodynamic motivation for dark energy was given in \cite{thermdarkenergy}.  Thermostatistical geometry, which incorporates configuration fluctuation distributions into the structure $(M, \langle g \rangle, P_R)$ of the expected geometry, is then introduced as the construct which contains the most we can know about a quantum spacetime's configuration.    
\end{sloppypar}
\begin{sloppypar}
Section \ref{expansion} addresses the problem of matter's zero-point energy by pointing out that $\langle G_{\mu\nu} \rangle+\Lambda \langle g_{\mu\nu}\rangle=0$ came from energy-shift invariant entropy and that coupling quantum spacetime's expected configuration with matter must be done with matter's expected configuration in an energy-shift invariant manner.  A brief mention is made of the mismatch between coupling geometry's expectation-valued entropy with matter's fundamental-field action which may be related to dark matter phenomenon.  Finally, it is noted that vacuum's curvature fluctuation entropy content could provide a direct physical picture of black hole entropy \cite{bhentropy} along with a mechanism to account for black hole information loss \cite{hawkingradiation}, \cite{BHinfoloss}.                   
\end{sloppypar}
\section{Many-body interaction equilibrium}\label{manybody}
\begin{sloppypar}
The introduction of statistical physics into the realm of quantum gravity provides a way around some of its notorious difficulties.  The essence of these difficulties, which will persist beyond issues of renormalizability, resides in the quantization of the background geometry itself.  Assuming we were doing this correctly and in a completely background-independent manner, the structure of quantum dynamics generically causes diffusion from a localized state in any free quantum system's dynamical variables.  Quantized geometry's small-scale configuration variables would then obey probability distributions with time-increasing variances and curvature fluctuation correlation lengths would be time-dependent resulting in a non-homogeneous spacetime.   At the root of this problem is the fact that we are quantizing a macroscopic many-body system without taking into account the tendency of all macroscopic many-body systems to configure at maximum entropy.  In light of this, the most natural and straightforward explanation of homogenous small-scale quantum structure is this tendency to configure at maximum entropy:  The many-body collection of local curvature fluctuations are in equilibrium with each other and distributed at maximum entropy.                   
\end{sloppypar}
\begin{sloppypar}
While the large-scale structure of the universe exhibits homogeneity which can be expressed as isometries of the large scale metric, how do we express the homogeneity of small-scale quantum structure?  If we repeatedly examined a sufficiently small neighborhood of quantum spacetime it would appear to be, \textsl{on average}, locally flat:  The small-scale curvature fluctuations have an expectation value of zero.  The key to understanding small-scale quantum structure resides in the curvature fluctuation probability distributions which must be identical for all small neighborhoods otherwise the fabric of our geometry would fail to be homogeneous.  This type of statistical homogeneity appears in the equilibrium ideal gas whose individual particle's momentum can be written 
\begin{equation}\label{idealgas}
p=\langle p \rangle+\varepsilon
\end{equation} 
The momentum fluctuations $\varepsilon $ (which are by no means small) about the macroscopic and emergent $\langle p \rangle$ obey probability distributions $P_{\varepsilon}$ which have no explicit positional dependence
\begin{equation}\label{probgas}
\partial_{\mu} P_{\varepsilon}=0
\end{equation}
which is equivalently expressed in terms of $\varepsilon$'s moments $\sigma_{\varepsilon\ell} \equiv \langle (\varepsilon-\langle \varepsilon \rangle)^{\ell} \rangle$ as
\begin{equation}\label{probepsilon}
\partial_{\mu} \sigma_{\varepsilon\ell}=0
\end{equation}
While a single, free quantum particle's momentum exihibits diffusion $\partial_t \sigma_{\varepsilon\ell}>0$ from a localized initial state, many-body interactions prevent quantum diffusion out of the ideal gas momentum distribution.        
\end{sloppypar}
\begin{sloppypar}
Quantum spacetime's statistically homogeneous small-scale structure can also be expressed as conditions on probability distributions when we write the quantum metric\footnote{Defined here as the result of a measurement and not an operator.} $g$ as fluctuations about a macroscopic and emergent $\langle g \rangle$:   
\begin{equation}\label{metricfluc}
g=\langle g \rangle+\delta g
\end{equation}
where the quantum configuration fluctuations $\delta g$ are also by no means ``small' and the precise meaning of $\langle g \rangle$ is given in the following definition.  Writing the quantum Riemann\footnote{Also defined as the result of a measurement and not an operator.} as fluctuations about its expectation\footnote{Expectation is defined as an average over an ensemble of configuration measurements.} value:
\begin{equation}\label{gbardef}
R_{\mu\nu\alpha\beta}=\langle R_{\mu\nu\alpha\beta} \rangle+\varepsilon_{\mu\nu\alpha\beta}
\end{equation}
The geometry $\langle g \rangle$ is then \textsl{defined} as a geometry with Riemann $\langle R_{\mu\nu\alpha\beta} \rangle$ while $\delta g$ is a configuration fluctuation with curvature fluctuation $\varepsilon_{\mu\nu\alpha\beta}$.  Under this definition, ensemble averaging the quantum Riemann removes the small-scale curvature fluctuations so that the expected local ``shape'' of $g$ equals the actual local ``shape'' of $\langle g \rangle$.  Contraction yields:
\begin{equation}\label{ricciFluc}
R=\langle R \rangle+\varepsilon 
\end{equation}
Sections \ref{gravex} and \ref{curvestat} demonstrate that if spacetime volume is locally created and destroyed in discreet units by quantum field operators, Ricci scalar fluctuations with probability distribution $P_\varepsilon$ will occur.  In order to preserve the homogeneity of vacuum's small-scale quantum structure, $P_{\varepsilon}$ cannot have explicit positional dependence 
\begin{equation}\label{probmetric}
\partial_{\mu} P_{\varepsilon}=0
\end{equation}
which is equivalently expressed in terms of $\varepsilon$'s moments $\sigma_{\varepsilon\ell} \equiv \langle (\varepsilon-\langle \varepsilon \rangle)^{\ell} \rangle$ as 
\begin{equation}\label{probsigma}
\partial_{\mu} \sigma_{\varepsilon\ell}=0
\end{equation}
Although $\sigma_{\varepsilon\ell}$ may have implicit positional dependence through fields, it is the lack of explicit dependence which keeps vacuum ``everywhere looking the same''.  Because quantum dynamics generically causes diffusion $\partial_t \sigma_{\varepsilon\ell}>0$ from localized initial states, the lack of diffusion in equations \eqref{probmetric} and \eqref{probsigma} should capture our attention.  Since the collection of local curvature fluctuations are a macroscopic, many-body system and quantum spacetime's neighbors can exchange gravitons with one another via non-linear field equations, the natural explanation for equations \eqref{probmetric} and \eqref{probsigma} is many-body interaction equilibrium.  Graviton exchange prevents quantum diffusion out of the entropy-maximizing fluctuation distribution $P_\varepsilon$ and this maintains the homogeneity of small-scale quantum structure.
\end{sloppypar}
\begin{sloppypar}
Because small-scale curvature fluctuations obey the laws of statistical physics, $\partial^2 g$ fluctuations will be exponentially suppressed by the familiar Boltzmann factor.  Consequently, extremely large fluctuations are highly improbable which is analogous to us not expecting to see a mole of gas spontaneously transfer all of its internal kinetic energy to one extremely energetic molecule.  Equilibrium along with entropy maximization enforces the smoothening of quantum geometry's small-scale configuration which justifies the use of Riemann normal coordinates in small-scale calculations.  The Riemann normal metric expansion allows for a more precise interpretation of equation \eqref{gbardef} which contains objects from two different geometries.  After identifying a point from $g$ with a point from $\langle g \rangle$, perform a metric expansion in each geometry from this common pole using the same set of initial geodesic directions in each expansion.  Equation \eqref{gbardef} at the pole now makes sense because it is comparing configurations from a common basis and all objects can be contracted\footnote{The $(+,-,-,-)$ convention is used throughout the paper.} at the pole with $\eta^{\mu\nu}$.  The expectation $\langle R^{\mu\nu\alpha\beta}_{pole} \rangle$ is defined in this basis as an average over an ensemble of configuration measurments\footnote{The physical process of configuration measurement is addressed in section \ref{curvestat}.}.
\end{sloppypar}
\begin{sloppypar}
The uncertainty principle's implication of extremely large curvature fluctuations \cite{kiefer} occurring almost everywhere neglects the laws of statistical physics, is characteristic of an infinite (canonical) temperature many-body system, and is unlikely to be the case with quantum spacetime.  The many-body interactions of grand canonical graviton exchange, in addition to enforcing $\partial_{\mu} \sigma_{\varepsilon\ell}=0$, act as the ``environment'' keeping the quantum metric ``localized'' providing an explicit source of decoherence.  
\end{sloppypar}
\section{The Riemann tensor's entropy}\label{riemphase}
\begin{sloppypar}
Quantifying the entropy of equilibriated local curvature fluctuations begins with an explicit construction of the Riemann tensor's phase space.  Because the Riemann obeys canonical statistics (explicitly constructed in section \ref{curvestat}), $\partial^2 g$ fluctuations are exponentially suppressed which keeps quantum spacetime locally smooth at the smallest scale.  This justifies the use of a Riemann normal metric expansion whose coefficients are now \textsl{randomly distributed variables}:
\end{sloppypar}
\begin{equation}\label{metricexpansion}
g_{\mu\nu}=\eta_{\mu\nu}+\frac{1}{3}\left[R_{\alpha\mu\nu\beta}\right]_{\xi=0}\xi^\alpha\xi^\beta+O\left(\xi^3\right)
\end{equation}
The Riemann tensor\footnote{The convention $R^a_{b c d}=\partial_c\Gamma^a_{b d}-\dotsb$ is being used.} at the pole is a linear combination of the random variables $\partial^2 g$ which map linearly into the curvature scalar $R$ which is an intrinsic random variable:
\begin{equation}\label{riematpole}
\left[R_{\mu\nu\alpha\beta}\right]_{\textsl{pole}}=\frac{1}{2}\left[\partial_\nu\partial_\alpha g_{\mu\beta}+\partial_\mu\partial_\beta g_{\nu\alpha}-\partial_\nu\partial_\beta g_{\mu\alpha}-\partial_\mu\partial_\alpha g_{\nu\beta}\right]_{\textsl{pole}}
\end{equation}
\begin{equation}\label{riccipole}
\eta^{\mu\beta}\eta^{\nu\alpha}\left[R_{\mu\nu\alpha\beta}\right]_{\textsl{pole}}=R_{\textsl{pole}}
\end{equation}
By choosing the same set of initial geodesic directions from the pole for each possible local configuration, the phase space $\Omega_{\textsl{Riem}}$ of $\left[R_{\mu\nu\alpha\beta}\right]_{\textsl{pole}}$ becomes a linear vector space in $\partial^2 g$ components.  After a linear transformation from $\partial^2 g$ into $\left(\mathbf{\ddot{g}}, R\right)$, where $\mathbf{\ddot{g}}$ labels Riemann's within phase space hyperplanes $\mathbb{S}_R$ of constant curvature scalar $R$, the degeneracies of the Riemann (due to its symmetries) reside within the hyperplanes $\mathbb{S}_R$ while two Riemann's differing in $R$ represent two different local configurations.  This representation allows the group of local diffeomorphisms at the pole $diff[M]_{pole}$ to be decomposed into diffeomorphisms within $\mathbb{S}_R$ and diffeomorphisms across $\mathbb{S}_R$:
\begin{equation}\label{diffdecomp}
diff[M]_{pole}=diff_{within}[M]_{pole}\otimes diff_{across}[M]_{pole}
\end{equation}
The action of $diff_{within}[M]_{pole}$ produces active as well as passive diffeomorphisms while the action of $diff_{across}[M]_{pole}$ produces only active diffeomorphisms.  Section \ref{curvestat} proves that the random variable $R$ is statistically independent from the random variables $\mathbf{\ddot{g}}$, therefore the probability of any state within $\Omega_{\textsl{Riem}}$ factorizes as: 
\begin{equation}\label{probOmega}
P(\mathbf{\ddot{g}}, R)=P(\mathbf{\ddot{g}})P(R)
\end{equation}
The entropy of $\left[R_{\mu\nu\alpha\beta}\right]_{\textsl{pole}}$ is then:
\begin{equation}\label{riemEntropy}
S=-k_B\sum_{\mathbf{\ddot{g}}}P\left(\mathbf{\ddot{g}}\right)\ln P\left(\mathbf{\ddot{g}}\right)-k_B\sum_R P\left(R\right)\ln P\left(R\right)
\end{equation}
\begin{sloppypar}
Normally, a Fadeev-Popov procedure would be used to eliminate the duplicate counting of passive states within the $\mathbf{\ddot{g}}$ sum.  However, the active elements within this sum must be discarded as well due to the underlying physical cause and structure of quantum spacetime's equilibrated local curvature fluctuations which will be explicitly developed in the following two sections.  Until then, a soft condensed-matter system with similar fluctuation structure is introduced in order to proceed.   
\end{sloppypar}
\begin{sloppypar}
Consider a flat, two dimensional, temperature-sensitive membrane which locally expands or contracts according to:
\end{sloppypar}
\begin{equation}\label{membrane}
\frac{A-\langle A \rangle}{\langle A \rangle}=k\bigl( T-\langle T \rangle \bigr)
\end{equation}
\begin{sloppypar}
As long as the membrane's temperature is maintained everywhere at $\langle T \rangle$, the membrane remains flat.  But if the temperature field spatially fluctuates about $\langle T \rangle$ with a normal distribution and position-independent variance, the local expansions and contractions produce local configuration fluctuations and a normally distributed\footnote{This is explicitly shown in equations \eqref{del}, \eqref{twodel}, and \eqref{PofRcurved}.} curvature scalar $R$ with position-independent variance.  Viewing the membrane differentially, temperature field gradients act as source terms for $2^{nd}$ order field equations which can, in principle, be solved for the membrane's configuration $g$.  Viewing the membrane as a Riemann surface, the membrane's physical configuration is determined by the temperature field's configuration which determines the location $(\mathbf{\ddot{g}}, R)$ of $\left[R_{\mu\nu\alpha\beta}\right]_{\textsl{pole}}$ within its phase space $\Omega_{\textsl{Riem}}$ (modulo passive diffeomorphisms).  Therefore, counting the states of $\mathbf{\ddot{g}}$ along with the states of $R$ over-counts the states of the membrane's local ``bumps''.   To avoid this, we associate the phase space of $T$ with the Ricci scalar's phase space $diff_{across}[M]_{pole}$ by setting $P(\mathbf{\ddot{g}})$ to a delta function.  
\end{sloppypar}
\begin{sloppypar}
The structure of these particular temperature fluctuations do not excite the membrane's expected macroscopic configuration $\langle g_{ij}\rangle=\delta_{ij}$ within the ``large'' phase space of $diff[M]$.  Section \ref{tsgeo} addresses the situation where $\langle g \rangle$ dynamically responds to the ``back reaction'' of fluctuation entropy maximization.  In this case, $\langle g \rangle$ obeys \textsl{Ehrenfest} equations of motion which again requires discarding the $\mathbf{\ddot{g}}$ sum in order to avoid over-counting.   In either case, the expected configuration $\langle g \rangle$ can be viewed as a macroscopic or thermodynamic variable which equilibrates within $diff[M]$.  
\end{sloppypar}
\begin{sloppypar}
Since $\Omega_{\textsl{Riem}}$ is a linear vector space in $\partial^2 g$ components, the Riemann's number of states per phase space volume is constant.  Since the transformation $\partial^2 g \mapsto (\mathbf{\ddot{g}}, R)$ is linear, the Ricci scalar's density of states $\rho_{\Omega}\equiv\ell^2_{\Omega}$ is also a constant constant allowing a straightforward entropy calculation using equation \eqref{riemEntropy} once $P(R)$ is known.    
\end{sloppypar}
\section{Grand canonical graviton exchange}\label{gravex}
\begin{sloppypar}
In order to find $P(R)$ for quantum spacetime, we need to know the effect quantum gravity has on small test volumes and to do this we consider the physical process of an expanding universe.  Because expansion is presumably governed by local quantum field theory, local expansion rates will fluctuate causing local $\frac{V-\langle V \rangle}{\langle V \rangle}$ fluctuations which must be in equilibrium with each other.  From the uniqueness arguments of the previous section's temperature-sensitive membrane, local $\frac{V-\langle V \rangle}{\langle V \rangle}$ fluctuations determine spacetime's $g=\langle g \rangle+\delta g$ fluctuations.  Finding the probability distribution satisfied by $\frac{V-\langle V \rangle}{\langle V \rangle}$, which reveals $P(R)$, can be accomplished if we consider expansion as the consequence of local quantum field operators which create (and destroy) vacuum.  Since nature exhibits sets of particles which are created (and destroyed) in discreet quanta, vacuum should also be created (and destroyed) in discreet quanta.  
\begin{equation}\label{vacuum}
\lbrace\varnothing\rbrace\rightleftarrows\lbrace\textsl{vacuum quanta}\rbrace
\end{equation}
Once created, however, vacuum quanta would have particle-like properties creating an ether.  The solution to this dilemma is the introduction of operators which create events of spacetime volume quanta 
\begin{equation}\label{stquanta}
\lbrace\varnothing\rbrace\rightleftarrows\lbrace\textsl{spacetime volume quanta}\rbrace
\end{equation}
which resembles a quantum field's virtual particle creation and destruction activity:
\begin{equation}\label{virtualpairs}
\lbrace\varnothing\rbrace\rightleftarrows\lbrace\textsl{virtual pairs}\rbrace
\end{equation}
\end{sloppypar}
\begin{sloppypar}
The number of volume quanta per spacetime volume $\rho=\frac{N}{V}$ must be finite otherwise the assumption of fundamental discreetness would be violated.  Because vacuum is Lorentz invariant, the operator's activity must be a statistical Poisson process with relative local positions that obey complete spatial-temporal randomness \cite{burton}.  Because $\rho$'s fluctuations must be in equilibrium, $\rho-\langle \rho \rangle=\frac{N-\langle N \rangle}{V}$ obeys grand canonical statistics.  In the ``system plus reservoir'' viewpoint of the grand canonical ensemble, the empty set $\{\varnothing\}$ on the LHS of equation \eqref{stquanta} plays the role of reservoir while the RHS plays the role of system.  Utilizing the alternative ``ensemble as reservoir'' viewpoint, we imagine stitching together a large collection of systems from the RHS of equation \eqref{stquanta} into a grand canonical reservoir.  In this viewpoint, the large collection of systems (volume quanta) becomes a spacetime geometry which acts as the reservoir for causally connected subsystems (small neighborhoods).  
\end{sloppypar}
\begin{sloppypar}
However, since the operators do not exist on a background but rather create the volume of the background itself, the ratio of quanta to spacetime volume does not fluctuate.  Because density fluctuations cannot occur, the Poisson-process operator fluctuations manifest as $\frac{V-\langle V \rangle}{\langle V \rangle}=\frac{N-\langle N \rangle}{\langle N \rangle}$ local volume fluctuations\footnote{This is most easily visualized in the ``system plus reservoir'' viewpoint.} which obey grand canonical statistics\footnote{Quantum spacetime behaves statistically as an incompressible boson event gas.} and cause local configuration fluctuations.  To find the statistics obeyed by $\frac{V-\langle V \rangle}{\langle V \rangle}$, we need a background-independent generalization of the grand canonical ensemble where instead of a system exchanging particles with a reservoir in a fixed background, the system is exchanging volume quanta with a volume reservoir causing local configuration fluctuations of the background itself.  
\end{sloppypar}
\begin{sloppypar}
To exhibit this generalization, we start with the fixed-background grand canonical ensemble which is set up in the ``system plus reservoir'' viewpoint as a small system of volume $V$ containing $N$ particles in contact with a large reservoir of volume $\tilde{V}\gg V$ containing a large number of particles $\tilde{N} \gg N$.  Particle exchange causes system density fluctuations at constant volume $\delta_V \rho$ where $\rho$'s distribution is governed by a chemical potential.  Equilibrium and maximum entropy are established when the chemical potentials of both system and reservoir are equal.  To make this set up background-independent, we pull the fixed background out from under the system plus reservoir while holding the particle density everywhere constant.  Volume quanta exchange causes system volume fluctuations at constant density $\delta_\rho V$ where $V$'s distribution is governed by a chemical potential.  Equilibrium and maximum entropy are established when the chemical potentials of both system and reservoir are equal. 
\end{sloppypar}
\begin{sloppypar}
If we now imagine stitching together a large number of these constant-density, background-free systems into one large volume (utilizing the ``ensemble as reservoir'' viewpoint), as individual systems exchange volume quanta with their neighbors, or equivalently \textsl{fluctuate in size under the action of operators}\footnote{This is the effect quantum gravity has on test volumes.} which locally create and destroy volume quanta, the large volume's geometry becomes locally bumpy.  In the limit of high quanta or operator density, the collection of volume quanta can be modeled as a smooth manifold.  In this limit, $\frac{V-\langle V \rangle}{\langle V \rangle}$ becomes a continuous Gaussian random variable which induces Gaussian local configuration fluctuations whose variance, chemical potential, or ``temperature'' cannot explicitly depend on position.  
\end{sloppypar}
\begin{sloppypar}
Equilibriated local configuration fluctuations have the dual description of equilibriated local graviton exchange.  A soft condensed-matter system's fluctuating membrane or surface \cite{membranephonons} has a similar dual description where equilibriated local curvature fluctuations can be viewed as equilibriated local phonon exchange.  From this viewpoint, the small-scale structure of quantum spacetime is dominated by grand canonical graviton exchange and its entropy content.   The conventional wisdom that gravitons cannot be in thermal or chemical equilibrium with each other since they interact very weakly with matter is incorrect.  This conclusion ignores both gravity's non-linear self-interaction and the tendency of macroscopic many-body systems to configure at maximum entropy.  A reservoir of gravitons will equilibrate directly with each other and do not require an intervening matter coupling.
\end{sloppypar}
\section{Curvature fluctuation statistics}\label{curvestat}
\begin{sloppypar}
A geodesic ball with fixed radius $\varepsilon$ centered on the pole of a Riemann normal coordinate system has volume which deviates, because of curvature fluctuations, from its expected volume according to
\begin{equation}\label{del}
\Delta \equiv \frac{V- \langle V \rangle}{\langle V \rangle}=\frac{1}{6(D+2)}\bigl(\langle R \rangle-R\bigr)\varepsilon^2+O(\varepsilon^3)
\end{equation}      
The randomly distributed $\Delta$ on the LHS of equation \eqref{del} corresponds with the ``system plus reservoir'' viewpoint of local volume fluctuations while the randomly distributed coefficients on the RHS correspond with the ``ensemble as reservoir'' viewpoint of local configuration fluctuations.  Operating with $\frac{\partial^2}{\partial \varepsilon^2 }\rvert_{\varepsilon=0}$ yields 
\begin{equation}\label{twodel}
\frac{\partial^2 \Delta}{\partial \varepsilon^2 }\Bigr\rvert_{\varepsilon=0}=\frac{1}{3(D+2)}\bigl(\langle R \rangle-R\bigr)
\end{equation}
Equilibriated local $\Delta$ fluctuations produce equilibriated local $R$ fluctuations whose variance, chemical potential, or ``temperature'' must be uniform in order to guarantee the homogeneity of small-scale quantum structure.  By the continuity theorem of transforms of normal distributions, the normally distributed $\Delta$ results in a normally distributed $R$ 
\begin{equation}\label{PofRcurved}
P(R)dR=\ell^2(2\pi)^{-\frac{1}{2}}e^{-\frac{\ell^4\left(R-\langle R \rangle\right)^2}{2}}dR
\end{equation} 
where the reciprocal variance $\ell^2=\frac{1}{\sigma}$ represents an invariant correlation length.  A key property (proven in this section) of the Ricci scalar's variance is its dependence on the expected Ricci scalar
\begin{equation}\label{sigmaR}
\sigma=\sigma_0 e^{\frac{L_G^2 \langle R \rangle}{4}}
\end{equation}
where $\sigma_0$ is the variance when $\langle R \rangle=0$, the parameter $L_G$  is a $D$-dependent universal constant, and the factor $1/4^{th}$ has been inserted for later convenience.
\end{sloppypar}
\begin{sloppypar}
Proving equation \eqref{sigmaR} begins with a discussion on how we ``adiabatically'' deform the expected macroscopic configuration of a quantum spacetime which possesses an intrinsic small-scale structure and scale.  An adiabatic deformation is defined as a diffeomorphism which preserves the operator density $\rho$, the correlation length $\ell_0 = \frac{1}{\sqrt{\sigma_0}}$\,, and the parameter $L_G$.  Locally ``stretching'' quantum spacetime without commensurately adding volume changes its intrinsic properties (causing an inhomogeneity) just as stretching a material solid changes its intrinsic properties.  Consequently, conformal transformations are non-physical since they alter $\rho$ and $\ell_0$ which alters the small-scale quantum structure and scale.  The only way nature can physically increase or decrease volume is to locally create or destroy volume and since this is done in a background-free manner, it preserves vacuum's small-scale quantum structure.
\end{sloppypar}
\begin{sloppypar}
Performing an adiabatic deformation from $\langle[R^{\mu\nu\alpha\beta}_1]\rangle$ into $\langle[R^{\mu\nu\alpha\beta}_2]\rangle$ in the neighborhood of a geodesic ball centered on the pole while holding the ball's expected volume constant sends the ball's radius $\varepsilon_1$ into $\varepsilon_2$, the curvature scalar $\langle R_1 \rangle$ into $\langle R_2 \rangle$, and the volume deviation $\Delta_1$ into $\Delta_2$:
\begin{subequations}\label{deformvolumes}
\begin{align}
\Delta_1 \equiv \frac{V_1-\langle V\rangle}{\langle V \rangle}=\frac{1}{6(D+2)}\bigl(\langle R_1 \rangle-R_1\bigr)\varepsilon^2_1+O(\varepsilon^3_1)\label{a}\\
\Delta_2 \equiv \frac{V_2-\langle V \rangle}{\langle V \rangle}=\frac{1}{6(D+2)}\bigl(\langle R _2 \rangle-R_2\bigr)\varepsilon^2_2+O(\varepsilon^3_2)\label{b} 
\end{align}
\end{subequations}
Since the ball's expected volume $\langle V \rangle$ was held constant during the adiabatic deformation, the LHS of equations \eqref{a} and \eqref{b} obey the same normal distribution.  Since $\langle R_1 \rangle \neq \langle R_2 \rangle$ implies $\varepsilon_1 \neq \varepsilon_2$, the Ricci scalar's variance $\sigma_1$ differs from $\sigma_2$ which proves the variance of $R$ is a function of expected configuration in the neighborhood of the pole.  
\end{sloppypar}
\begin{sloppypar}
To calculate the variance of $\frac{\partial^2 \Delta}{\partial \varepsilon^2 }\bigr\rvert_{\varepsilon=0}$, we measure $\Delta$ as a function of radius near the pole for one ensemble member, perform $\frac{\partial^2}{\partial \varepsilon^2 }\bigr\rvert_{\varepsilon=0}$ on this member's volume data, repeat over ensemble members, and from this ensemble data calculate the variance.  The physical process of volume measurement introduces additional quantum mechanical uncertainties and, because of matter's interaction with geometry, produces a small change $\langle R \rangle+ \langle \delta R \rangle$ in equation \eqref{sigmaR}.  However, since we are summing over an ensemble of measurements, the central limit theorem guarantees that the Gaussian structure of the Ricci scalar's distribution remains unchanged.  This measurement-induced rescaling of the ``bare'' $\sigma$ will be absorbed into a free parameter in the next section.  Because we performed $\frac{\partial^2}{\partial \varepsilon^2 }\bigr\rvert_{\varepsilon=0}$ on our data, $\sigma$ does not receive contributions from $\langle g \rangle$ derivatives higher than $2^{nd}$ order.  Additionally, the structure of the Riemann normal metric expansion and the geodesic ball volume expansion rules out contributions from a lower-order Chern-Simons term.  The Ricci scalar's variance therefore depends on expected configuration only through $\langle R \rangle$:
\begin{equation}\label{W}
\sigma=W\negthinspace\left(\langle R \rangle \right)\sigma_0
\end{equation}
\end{sloppypar}
\begin{sloppypar}
Next we note that given a fixed-background grand canonical ensemble, the variance of $\frac{N-\langle N \rangle}{\langle N \rangle}$ increases with decreasing sample size $\langle N \rangle$.  By holding density fixed, the variance of $\frac{V-\langle V \rangle}{\langle V \rangle}$ also increases with decreasing sample size $\langle V \rangle$ in the background-free graviton exchange ensemble where we change the fixed-radius geodesic ball's sample size $\langle V \rangle$ via adiabatic deformation.  Increasing $\langle R \rangle$ decreases $\langle V \rangle$ which increases the variance of $\Delta$, therefore the Ricci scalar's variance $\sigma$ increases with increasing $\langle R \rangle$ as well.  To see the form of this dependence, we note that adiabatically deforming an $\langle R \rangle=0$ configuration in the neighborhood of a fixed-radius geodesic ball into symmetrical $\pm \langle \delta R \rangle$ deformations about $\langle R \rangle=0$ results in symmetrical $2^{nd}$ order $\mp \langle \delta V \rangle$ expected volume changes.  This symmetry implies $W$ must satisfy 
\begin{equation}\label{Wsymmetry}
W\negthinspace\left(+\langle R \rangle\right)W\negthinspace\left(-\langle R \rangle\right)=1
\end{equation}
Since $W\negthinspace\left(0\right)=1$, $W$ must be of the form
\begin{equation}\label{W}
W=e^{\frac{L_G^2\langle R \rangle}{4}}
\end{equation}
Which proves equation equation \eqref{sigmaR}.  Since $P(R)$ is independent of the Riemann's location $\mathbf{\ddot{g}}$ within phase-space hyperplanes $\mathbb{S}_R$, the phase space variables $\mathbf{\ddot{g}}$ and $R$ are statistically independent which was asserted in equation \eqref{probOmega}. 
\end{sloppypar}
\section{Thermostatistical geometry}\label{tsgeo}
\begin{sloppypar}
The fluctuation entropy of a single system, which in this case is a single spacetime point, is found using the distribution and variance of equations \eqref{PofRcurved} and \eqref{sigmaR} respectively yielding   
\begin{equation}\label{rhoS}
\rho_S=\frac{1}{2}k_B\ln\left(\frac{2\pi e}{\ell^4}\right)
\end{equation}
Inserting the Ricci scalar's density of states $\rho_\Omega\equiv \ell_\Omega^2$, which can be done using dimensional analysis or through a change of variables in equation \eqref{PofRcurved}, inserting $\ell^4$ from equation \eqref{sigmaR}, and then multiplying by the number of systems per spacetime volume $\rho_0$ (the ``volume graininess'' of quantum spacetime) yields quantum spacetime's Einstein-Hilbert entropy per volume:
\begin{equation}\label{flucentropy}
\rho_S=\frac{1}{2}k_B\rho_0\left[\ln\left(\frac{2\pi e\cdot \ell^4_\Omega}{\ell_0^4}\right)+\frac{1}{2}L_G^2\langle R \rangle\right]
\end{equation}
Bringing the coefficient of $\langle R \rangle$ to 1 yields:
\begin{equation}\label{expectedEH}
\langle R \rangle+2\Lambda=\frac{4\rho_S}{k_B\rho_0L_G^2}
\end{equation}
\begin{equation}\label{lamda}
\Lambda=\frac{1}{L_G^2}\ln\left(\frac{2\pi e\cdot \ell_\Omega^4}{\ell_0^4}\right)
\end{equation}
The total Einstein-Hilbert entropy is proportional to the Einstein-Hilbert action \textsl{evaluated on the expected macroscopic geometry} $\langle g \rangle$ which is maximized when  
\begin{equation}\label{varS}
\delta \negthinspace\int\negmedspace d^D\!\langle x \rangle \sqrt{-\langle g \rangle}\,\bigl(\langle R \rangle+2\Lambda\bigr)=0
\end{equation}
where $\langle x \rangle$ is defined as a coordinate system with metric $\langle g \rangle$.
This produces the Ehrenfest equations of motion\footnote{This requires dropping the Riemann's $\mathbf {\ddot{g}}$ entropy sum which was asserted in section 3.} for $\langle g \rangle$ which are identical to Einstein's \textsl{expectation-valued} field equations: 
\begin{equation}\label{MFE}
\langle G_{\mu\nu} \rangle+\Lambda \langle g_{\mu\nu} \rangle=0
\end{equation}
Contracting equation \eqref{MFE} and using equation \eqref{expectedEH} yields 
\begin{equation}\label{rholambda}
\Lambda=\frac{\langle R \rangle}{D}=\frac{4\rho_S}{(D+2)k_B\rho_0L_G^2}
\end{equation} 
\end{sloppypar}
\begin{sloppypar}
The cosmological constant $\Lambda$  as expressed in equation \eqref{lamda} depends on three parameters two of which are in a ratio.  The parameter $L_G$ depends on $D$ alone and can in principle be determined from methods of probability which leaves only one free parameter $\frac{\ell_\Omega^4}{\ell_0^4}$.  The previous section's measurement-induced rescaling of the Ricci scalar's variance $\sigma$ can be absorbed into this parameter.
\end{sloppypar}
\begin{sloppypar}
The change of viewpoint, from minimization of action to maximization of entropy, is necessary because quantum spacetime is an isolated, macroscopic, many-body system evolving at fixed energy.  It is interesting to note that $\langle g \rangle$ has been \textsl{dynamically determined} via entropy maximization from the ``back reaction'' of $\delta g$ fluctuations about $\langle g \rangle$ and that this formulation is completely background-independent. 
\end{sloppypar}
\begin{sloppypar}
Just as an ideal gas locally exchanges energy quanta which maintains maximum entropy and a uniform temperature, quantum spacetime\footnote{Quantum spacetime behaves statistically as an incompressible boson event gas.} locally exchanges curvature quanta which maintains maximum entropy and a uniform Ricci scalar ``temperature''.  Although the macroscopic expected configuration $\langle g \rangle$ is dynamic, it is the local fluctuations $\delta R$ about $\langle R \rangle$ which remain in equilibrium with each other and it is their probability distributions which have no explicit dependence on position.
\end{sloppypar}
\begin{sloppypar}
Because small-scale quantum spacetime experiences random spatial-temporal fluctuations and many-body interactions, its exact configuration $g$ can never be known.  The most we can know is the expected configuration $\langle g \rangle$ and the distribution of curvature fluctuations about $\langle g \rangle$.  This provides the same information as a quantum system's wavefunction which generates the expectation values of observables along with all moments about those expectation values.  Because the classic Einstein-Hilbert action, apart from dark energy and dark matter phenomenon, accurately describes our large scale $\langle g \rangle$, its structural identity with the expectation-valued Einstein-Hilbert entropy suggests it is not, in fact, an action.  From the viewpoint of statistical physics, the Hamiltonian $\left(R-\langle R \rangle\right)^2$ along with the Ricci scalar's ``temperature'' $\sigma^2=\sigma_0^2 e^{\frac{L_G^2 \langle R \rangle}{2}}$ statistically results in macroscopic deSitter spacetime absent matter fields.  As an aside, the ``zero'' of any quantum gravity's Hamiltonian, which in this case is $\langle R \rangle$, must be dynamically determined in order to achieve background independence.          
\end{sloppypar}
\begin{sloppypar}
Thermostatistical (TS) geometry incorporates the probability distribution of equilibriated local curvature fluctuations into the structure of the macroscopic expected geometry:
\begin{equation}\label{TS}
\left(M,\langle g \rangle\right)\xrightarrow{\delta R}\left(M, g\right)\xrightarrow{\textsl{TS}}\left(M,\langle g \rangle, P_R\right)
\end{equation}
\end{sloppypar}
The distribution $P_R$ can alternatively be written as the entropy field $\rho_S$ or, when the distribution is normal, as a ``temperature'', variance, or correlation length.  In the deterministic case where no fluctuations are present, $P_R$ becomes a delta function, the entropy, ``temperature'' and variance become zero, and the correlation length becomes infinite.  Since $(M, \langle g \rangle)$ can be any Riemannian geometry, TS geometry contains all of deterministic Riemannian geometry as the zero-entropy special case.  A soft condensed-matter system's fluid membrane provides a physical example of a TS geometry with temperature-dependent local curvature fluctuations and correlation lengths that diverge at low temperature \cite{membranecorrelation}.  TS geometry's expected configuration is completely determined by the canonical probability $P_R$ which differs significantly from action-based dynamics.
\begin{sloppypar}
TS geometry inherits several intrinsic length scales.  The correlation length $\ell$ is an invariant under passive adiabatic deformations and plays the role of a ``curvature corrected'' Planck length.  The correlation length $\ell_0=\ell\negthinspace\cdot\negthinspace e^{\frac{L_G^2 \langle R \rangle}{8}}$ is a topological invariant under active adiabatic diffeomorphisms and plays the role of an ``absolute'' Planck length.  The Ricci scalar's density of states $\ell_\Omega^2$ and the $D$ dependent parameter $L_G$ are both universal constants.  All of these lengths are the consequence of statistical physics and do not require a modification of special relativity to explain their invariance \cite{DSR1}, \cite{DSR2}.    
\end{sloppypar}
\section{Entropy-driven expansion}\label{expansion}
\begin{sloppypar}
Vacuum has entropy content, quantum spacetime has the ability to create more vacuum, more vacuum means more entropy, thus expansion is simply the consequence of a many-body system occupying its available phase space.  Fluid membranes, which in addition to undergoing local fluctuations have the ability to change their size via molecular exchange with their surrounding medium, posses dynamics similar to spacetime geometry \cite{membranelambda}.  Their effective Hamiltonian's ``cosmological constant'' term can be viewed as the chemical potential of membrane creation, the membrane's surface tension, or as the consequence of entropy maximization.  
\end{sloppypar}
\begin{sloppypar}
The pressure of quantum spacetime's small-scale curvature fluctuations causes expansion which differs from Weinberg's \cite{weinberg} classifications (symmetry, anthropic, tuning, modification, and quantum gravity) all of which rely on action.  Entropy-based expansion also differs from newer classifications \cite{DEreview} (holography, back-reaction, and phenomenological models) which also rely on action.  Of these, the back-reaction from sub-Hubble scale fluctuations is logically related to entropy-based expansion.  If Planck-scale fluctuations had been considered, the requirement of small-scale statistical homogeneity necessitating the introduction equilibrium would have led directly to entropy and its maximization.
\end{sloppypar}
\begin{sloppypar}
Although the Ehrenfest equations of motion for $\langle g \rangle$ are identical to Einstein's expectation-valued field equations, they were derived from entropy which is invariant under global energy shifts.  Consequently, coupling matter to $\langle g \rangle$ can only be done using matter's expectation values in a manner which is also invariant under global energy shifts.  This raises the possibility that quantum matter's zero-point energy problem may not be a problem after all because this result is based on action, not entropy, and is not energy shift invariant.  From the viewpoint of a perfect fluid, the coincidence of \ $pressure=-\rho_{energy}$ occurs because it is not a perfect fluid behind the $\Lambda$ term, it is local curvature fluctuation entropy.  Because classic gravitation is essentially coupling matter's fundamental-field action with quantum spacetime's expectation-valued entropy, discrepancies (such as dark matter phenomenon) should be expected from extended many-body gravitating systems.  These ideas are currently being investigated and will be addressed in a future publication.    
\end{sloppypar}
\begin{sloppypar}
Since quantum spacetime's curvature fluctuations have entropy per volume proportional to $\langle R \rangle+2\Lambda$, a direct physical picture of black hole entropy along with a mechanism to account for black hole information loss may now be possible.  These ideas are also being investigated and will be addressed in a future publication.  The increasing entropy accompanying an expanding universe buttresses the arrow of time role played by the $2^{nd}$ law of thermodynamics.           
\end{sloppypar}
\section{Conclusion}\label{conclusion}
\begin{sloppypar}
The small-scale structure of quantum spacetime is dominated by local curvature fluctuations whose many-body characteristics necessitate a statistical approach to their description.   Because the small-scale structure is statistically homogeneous, curvature fluctuation probability distributions cannot have explicit positional dependence which is the signature of equilibrium.  Local curvature fluctuations obey grand canonical statistics which exponentially suppress large fluctuations keeping quantum spacetime Riemannian at the smallest scale, prevents diffusion, and maintains decoherence.   Fluctuation entropy is proportional to the Einstein-Hilbert action evaluated on the macroscopic expected geometry and includes a small, positive cosmological constant.  Entropy maximization yields quantum spacetime's Ehrenfest equations of motion which are identical to Einstein's expectation-valued field equations.  This background-free formulation reveals curvature fluctuation entropy as the source of expansion and raises the possibility that matter's zero-point energy problem, which is action-based and not energy shift invariant, may not be a problem after all. 
\end{sloppypar}

\end{document}